\documentclass[preprint, amsmath, amssymb, aps, a4]{revtex4-2}
\usepackage{graphicx}
\usepackage{lineno}
\usepackage{bm}
\usepackage{txfonts}
\usepackage{hyperref}
\usepackage{caption}
\usepackage{amsmath}
\usepackage{amssymb}
\usepackage[section]{placeins}
\usepackage{tabularx}
\usepackage{braket}
\usepackage{multirow}
\usepackage{subcaption}
\usepackage{rotating}
\usepackage{xcolor}

\date{\today}
\newcommand{\be}{\begin{eqnarray}}
\newcommand{\ee}{\end{eqnarray}}

\newcommand{\bfk}{{\bf k}_{\perp}}

\begin{document}
\title{Radiative Transitions for the Ground and Excited Charmonia States}
\author{Anurag Yadav$^{1}$}
\email{anuragyadav1431905@gmail.com}
\author{Satyajit Puhan$^{1}$}
\email{puhansatyajit@gmail.com}
  \author{Harleen Dahiya$^{1}$}
\email{dahiyah@nitj.ac.in}
\affiliation{$^1$ Computational High Energy Physics Lab,  Department of Physics,  Dr.  B. R.  Ambedkar National
	Institute of Technology,  Jalandhar,  144008,  India}

\date{\today}%
\begin{abstract}
In this work, we have investigated the physical properties like decay constants, radiative transitions, decay widths, and branching ratios for the ground and radially excited charmonia states. For the numerical calculations, we have adopted the light-front quark model (LFQM). We have studied  $\chi_{c0}\rightarrow J/\psi+\gamma $ and $\psi(2S)\rightarrow\chi_{c0}+\gamma$,  $h_c(1P)\rightarrow\eta_c(1S)+\gamma $, and $\eta_c(2S)\rightarrow h_c(1P)+\gamma $ transitions in this work. We have also studied the $Q^2$ dependence of the electric dipole transition form factors (TFFs) for all four transitions and compared them with the available lattice QCD and basis light-front quantization (BLFQ) results. Using the TFFs results, we have calculated the decay widths and branching ratios for these transitions. The electric dipole matrix element is decomposed into its Lorentz structures and $E_1(Q^2)$ is extracted from a combination of the longitudinal and transverse plus-current amplitudes canceling the accompanying Coulomb form factor identically. The light-front spin-orbit wave functions of the $P$-wave states $\chi_{c0}$ ($^3P_0$) and $h_c$ ($^1P_1$) are constructed from the Melosh transformation of the corresponding $|(LS)JJ_z\rangle$ states, rather than from the $S$-wave pseudo-scalar and vector ones. The charm quark mass and the harmonic scale are fixed by a variational fit to the two $1S$ masses alone. Our numerical results of decay constants,  decay widths, and branching ratios are overall in good agreement with available experimental, theoretical and lattice simulation data.

\vspace{0.1cm}
\end{abstract}
%
\maketitle
%
%

\section{Introduction\label{secintro}}
The study of charmonium, a bound system of a charm quark and its antiquark, continues to be a cornerstone in understanding strong interactions. Being one of the cleanest environments for probing Quantum Chromodynamics (QCD) \cite{Brodsky:1997de, Zhang:1997dd}, charmonium spectroscopy offers a unique window into both the perturbative and non-perturbative regimes. Among the various processes that connect different charmonium states, the electric dipole (E1) transitions are of particular importance. These radiative transitions, governed by well-defined selection rules, provide critical information about the internal structure of quarkonia, the nature of the QCD potential, and relativistic effects in heavy quark systems. Charmonium occupies a valuable intermediate position within QCD,  being neither in the purely non-relativistic regime nor in the regime where chiral symmetry breaking dominates.  This makes it a relatively clean
system to study non-perturbative QCD within the quark models. 

\par The ground state vector charmonia $J/\psi$ was first discovered in 1974 by Brookhaven National Lab (BNL) \cite{PhysRevLett.33.1404} and Stanford Linear Accelerator Center (SLAC) \cite{PhysRevLett.33.1406} having a charm quark and its antiquark. As expected, it was discovered to have a close resemblance to positronium with a range of resonances that correspond to different heavy quark pair excitations. It was anticipated that the charmonium would contribute to the study of hadronic dynamics in the same way as the hydrogen atom does to atomic physics \cite{NOVIKOV19781}. Similarly, after the discovery of $X(3872)$ charmonium state at Belle in 2003 \cite{Belle:2003nnu},  it has been an interesting topic to understand the transitions occurring in between the charmonia states. In the BaBar  and Belle  B-factories, numerous charmonia and charmonia-like states were found in the first ten years of this century \cite{BaBar:2001yhh,Belle:2000cnh,BaBar:2014omp}. For a better understanding of confinement in the higher excited state of charmonia, the study of mass spectra, decay constants,  decay widths, branching ratios, and transition form factors (TFFs) becomes important. These observables provide deep insight and information about the charmonia and their decay processes.

The TFFs describe the structure-dependent effects in the interaction between a hadron and a virtual photon with momentum transfer $Q^2$. It encodes the spatial and dynamical properties of the hadron and is crucial for understanding its internal structure. These form factors encapsulate the effects of strong interaction in electromagnetic or weak transitions between different charmonium states. Beijing Spectrometer III (BES-III) is one of the leading experiment to study the radiative transitions among the charmonia decays \cite{BES:2012uhz,BES:2006rsb,BESIII:2020nbj}. However, there are a very few theoretical works reported for the radially excited charmonia decays compared to the ground state charmonia decays. The decay width of a hadron represents the probability per unit time that the particle will decay into specific final states and is mainly dependent on the TFF of the particular decay process in which the hadron decays.  

Electromagnetic transitions among the charmonia \cite{Voloshin:2007dx} are classified in terms of electric and magnetic multipoles. The most essential ones are E1 (electric dipole) and M1 (magnetic dipole) transitions. Higher multipole modes like E2, E3, M1. . . .  transitions also exist, but they are suppressed. Further, the electric dipole transition happens more chronically as compared to the magnetic dipole transition and E1 transition has a significant contribution to the total decay width of charmonium. The main characteristic of an E1 transition is that there is no change in the spin angular momentum, but there is a change in the orbital angular momentum of the particle by one unit. As a result, the final state has a different parity and C-parity than the initial one.

\par On the theoretical side, the radiative transitions of charmonia have been studied in several frameworks, and each of them handles the relativistic and non-perturbative effects in its own way. The oldest are the potential models. In the non-relativistic (NR) potential model, the E1 width is fixed by the dipole matrix element between the radial wave functions of the initial and final states \cite{Eichten:1978tg,Eichten:2007qx}. The relativized Godfrey-Isgur (GI) model includes the leading relativistic corrections to the kinetic energy and to the potential \cite{Godfrey:1985xj,Barnes:2005pb}, and relativistic quark models of the same type have been used for the whole charmonium spectrum and its transitions \cite{Ebert:2002pp,PhysRevD.95.034026}. These models describe the low-lying transitions reasonably well, but for the radially excited states their predictions depend strongly on the form of the potential and on the way the relativistic corrections are included \cite{Barnes:2005pb,Eichten:2007qx}. A more systematic treatment is provided by non-relativistic QCD (NRQCD) \cite{Bodwin:1994jh} and its low-energy version, potential NRQCD (pNRQCD) \cite{Brambilla:2004jw}, where the transition amplitude is expanded in powers of the heavy-quark velocity $v$. For the E1 transitions, the relativistic corrections at order $v^2$ have been worked out in Ref. \cite{Brambilla:2012be}. For charmonium, however, $v^2\approx0.3$ and the expansion converges slowly, particularly for the $2S$ and $1P$ states \cite{brambilla2011heavy}.

\par The Bethe-Salpeter (BS) approach treats the $c\bar c$ pair as a relativistic bound state. It has been used for the decay constants of the $P$-wave charmonia \cite{Zhou:2020bnm} and, more recently, for the $\chi_{cJ}\rightarrow\psi\gamma$ transitions \cite{Pei:2024hzv}. Lattice QCD provides the first-principles results. The E1 form factors of $\chi_{c0}\rightarrow J/\psi\gamma$ and $h_c\rightarrow\eta_c\gamma$ were first calculated in quenched lattice QCD by Dudek \textit{et al.} \cite{Dudek:2006ej}, and the transitions involving excited states, including $\psi(2S)\rightarrow\chi_{c0}\gamma$, were obtained later using the variational method \cite{Dudek:2009kk}. Further results are available from $N_f=2$ twisted-mass simulations \cite{Chen:2011kpa,Becirevic:2012dc,li2021radiative} and from the recent dynamical calculation of Ref. \cite{Delaney:2023fsc}. These results still carry sizeable systematic uncertainties from quenching, discretisation and the treatment of the charm quark on the lattice, and, as we will see in Sec. \ref{results}, different groups do not always agree with one another. On the light front, the BLFQ approach solves the $c\bar c$ Hamiltonian with a confining potential and one-gluon exchange \cite{Li:2015zda,Li:2017mlw}, and has been used for the M1 \cite{Li:2018uif} and E1 \cite{Wang:2023nhb} transition form factors of charmonia. The LFQM used in the present work is simpler than BLFQ. The wave function is not obtained by diagonalising a Hamiltonian but is taken as a Gaussian, with the parameters fixed variationally, and the M1 transitions of heavy mesons have been studied in this model earlier \cite{Choi:2007se}. Its advantage is that the TFF comes out as a direct overlap of the light-front wave functions and can be followed over a wide range of $Q^2$, so that it can be compared directly with the lattice and BLFQ results. For the E1 transitions involving the radially excited charmonia, such a study is still missing, and this is what we address in the present work.

\par In this work, we have studied the E$1$ transitions among the radially excited charmonia decays using the light-front quark model (LFQM) \cite{ Cheng:2003sm, cheng1997mesonic, Terentev:1976jk,Puhan:2023hio,Acharyya:2024enp,Acharyya:2024tql}. The LFQM is a non-perturbative approach framework describing the structure and dynamics of hadrons as well as the behavior of quarks within them. The nature of LFQM is gauge invariant and relativistic.  Further, LFQM has the benefit of concentrating on the hadrons' valence quarks, which are the main components to describe the general structure and characteristics of hadrons. In the strong interaction regime, hadrons can be described by the LFQM when perturbative QCD computations are unable to do so.  However, LFQM does not account for the effect of confinement and increased Fock-state contributions, which are crucial in describing the whole dynamics of hadrons.  The hadron structure is exclusively described by LFQM in the lower Fock-states. 

\par In this work, we have investigated the P-wave and S-wave charmonia decays  
$\chi_{c0}(1P)\rightarrow J/\psi(1S)+\gamma $, $\psi(2S)\rightarrow\chi_{c0}(1P)+\gamma$,  $h_c(1P)\rightarrow\eta_c(1S)+\gamma $, and $\eta_c(2S)\rightarrow h_c(1P)+\gamma $. It is interesting to mention that $\chi_{c0}(1P)\rightarrow J/\psi(1S)+\gamma $ transition is significant in understanding the P-wave charmonia as there is consensus  regarding the decay width of particle data group (PDG) \cite{ParticleDataGroup:2022pth}, CLEO data \cite{CLEO:2005efp}  and measured BESIII  value \cite{BESIII:2017gcu}. This discrepancy in the values of decay width requires a theoretical investigation of this transition for a more detailed understanding. For  
 $ \psi(2S)\rightarrow\chi_{c0}(1P)+\gamma$ and  $h_c(1P)\rightarrow\eta_c(1S)+\gamma$, where there is no experimental data available for, we have studied the TFF as well as decay width of these transitions and compared it with other theoretical model predictions. For the  $\eta_c(2S)\rightarrow h_c(1P)+\gamma $   transition, there is no experimental data available for decay width; we have calculated the decay width of this transition and compared it with theoretical data and also calculated the respective branching ratio. 

The paper is arranged as follows. In Sec. II, we have  discussed LFQM along with the spin and meson wave functions using the light-front formalism. The input parameters and the wave form of different mesons have been discussed in this section. The decay constants of the excited state mesons are calculated. Details pertaining to the transition form factor correlator and radiative decay width relation are also presented in this section. In Sec. III, we have discussed and  presented the numerical results of decay widths, branching ratios and transition form factors of all four transitions. In Sec. IV,  we have summarized our work.

 \section{Methodology}\label{satya}
 \subsection{Light-front quark model}
 LFQM provides an ideal framework for describing the mesons state in terms of their constituent quarks and antiquarks. The minimal hadron fock-state wave function based on light-front quantization in the form of quark-antiquark is expressed as \cite{Choi:1999nu,Dhiman:2019ddr,Dhiman:2017urn}
\begin{eqnarray}
\label{fockstate0}|M(P,  J,  J_z)\rangle &=& \sum_{\lambda_q,  \lambda_{\bar{q}}} \int \frac{\mathrm{d}x \,  \mathrm{d}^2 \mathbf{k}_{\perp}}{\sqrt{x(1-x)} 16 \pi^3} \,  \psi_{\lambda_q,  \lambda_{\bar{q}}}^{J J_z}(x,  \mathbf{k}_{\perp}) |x,  \mathbf{k}_{\perp},\lambda_q,\lambda_{\bar q} \rangle. 
\end{eqnarray}
Here, $|M(P, J, J_z)\rangle$ and  $P=(P^+, P^-, {P_{\perp}})$ are the eigenstate and four-vector total momentum of the meson, respectively. (${J,J_z}$) is the total angular momentum of the meson state.  $\mathbf{k}=(K^+, K^-, \mathbf{k}_\perp)$ is the four vector momenta of the active quark with $x=\frac{K^+}{P^+}$ being the longitudinal momentum fraction of the active quark. $1-x$ is the longitudinal momentum fraction carried by the antiquark. $\lambda_{ q, (\bar q)}$ is the helicity of the quark (antiquark). 
\par The light-front wave function (LFWFs) $\psi_{\lambda_q, \lambda_{\bar q}}^{J J_{z}}$  for mesons can be expressed as \cite{Arifi:2022pal} 
\begin{equation} \label{wave}
\psi^{J(J_{z})}_{\lambda_q, \lambda_{\bar q}}(x,\mathbf{k}_\perp) = \phi_{nS(nP)}{(x, {\mathbf{k}_\perp})}\chi^{JJ_{z}}_{\lambda_q, \lambda_{\bar q}}(x, {\mathbf{k}_\perp}), 
\end{equation}
where $\phi_{nS(nP)}$ is the radial (orbital) wave function, and $\chi^{JJ_{z}}_{\lambda_q, \lambda_,\bar q}$ is a spin-orbit wave function obtained by interaction-independent Melosh transformation. In this work, $nS (nP)$  is used for various radially (orbitally) excited states of mesons. Covariant form of the spin-orbit  wave functions for pseudo-scalar (J=0) and vector (J=1) mesons are respectively expressed as \cite{Arifi:2022pal,Acharyya:2024tql}
\begin{eqnarray}\label{fockstate}
\chi_{\lambda_q, \lambda_{\bar{q}}}^{00}
&=&-\frac{1}{\sqrt{2}\widetilde{M}}{{\bar{u}}_{\lambda_{\bar{q}}}}(\mathrm{p}_q){\gamma_{5}}{v_{\lambda_{\bar{q}}}}(\mathrm{p}{}_{\bar{q}}), 
\end{eqnarray}
and
\begin{equation} \label{vector}
\chi_{\lambda_q, \lambda_{\bar{q}}}^{1 J_z} = -\frac{1}{\sqrt{2}\widetilde{M}} \bar{u}_{\lambda_{\bar{q}}}(\mathrm{p}_q) \left[ \not\!\epsilon(J_z) - \frac{\epsilon \cdot (\mathrm{p}_q - \mathrm{p}_{\bar{q}})}{M_o + m_q + m_{\bar{q}}} \right] v_{\lambda_{\bar{q}}}(\mathrm{p}_{\bar{q}}),
\end{equation}
where $\widetilde{M}=\sqrt{{M_o}^2-(m_q-m_{\bar{q}})^2}$. Here, $M_o$ is the invariant meson mass square under the influence of boost with $m_{q(\bar q)}$ being the quark (antiquark) masses defined as
\begin{eqnarray}
    M_o^2=\frac{\mathbf{k}_\perp^2+m_q^2}{x}+\frac{\mathbf{k}_\perp^2+m_{\bar{q}}^2}{1-x}.
\end{eqnarray}
$\epsilon^{\mu}(J_z)=(\epsilon^+, \epsilon^-, \epsilon_{\perp})$ in Eq. (\ref{vector})
is the polarisation vector of the vector meson and is represented as 
\begin{equation}
   \epsilon^{\mu}(\pm1)=\left(0, \frac{2}{P^+}\epsilon_{\perp}(\pm). P_{\perp}, \epsilon_{\perp}(\pm)\right), 
\\
 \epsilon^{\mu}(0)=\frac{1}{M_o}\left(P^+, \frac{-{M_o}^2+P_{\perp}^2}{P^+}, P_{\perp}\right),    
\end{equation}
where
\begin{equation}
    \epsilon_{\perp}(\pm1)=\mp\frac{1}{\sqrt{2}}(1, \pm i).
\end{equation}
The $u$ and $v$ are the Dirac spinors in Eq. (\ref{vector}) with $\gamma$ being the Dirac matrices. The spin-orbit wave function obeys the unitary condition automatically as $\langle \chi^{JJ_{z}}_{\lambda_q, \lambda_,\bar q}| \chi^{JJ_{z}}_{\lambda_q, \lambda_,\bar q}\rangle=1$.

Equations (\ref{fockstate}) and (\ref{vector}) hold for the $S$-wave pseudo-scalar ($^1S_0$) and vector ($^3S_1$) mesons, where $L=0$ and the meson spin comes entirely from the quark spins. Since the E1 transitions studied here always involve a $P$-wave partner, either $\chi_{c0}(0^{++})$ or $h_c(1^{+-})$, therefore, in these cases, the orbital motion carries the angular momentum. In case Eqs. (\ref{fockstate}) or (\ref{vector}) are used for such a state, it would lead to a $S$-wave spin structure, which it does not have. We therefore build the spin-orbit wave function for a general $|(LS)JJ_z\rangle$ state from the interaction-independent Melosh transformation \cite{Melosh:1974cu,Jaus:1989au,Jaus:1991cy},
\begin{eqnarray}\label{melosh}
\mathcal{M}(m, \xi, \mathbf{k}_\perp)&=&\frac{(m+\xi M_o)+i\bm{\sigma}\cdot(\hat{n}\times\mathbf{k}_\perp)}{\sqrt{(m+\xi M_o)^2+\mathbf{k}_\perp^2}},
\end{eqnarray}
with $\hat n=(0,0,1)$, evaluated at $(\xi,\mathbf{k}_\perp)=(x,\mathbf{k}_\perp)$ for the quark and $(1-x,-\mathbf{k}_\perp)$ for the antiquark, so that \cite{Jaus:1991cy,Cheng:2003sm}
\begin{eqnarray}\label{chigeneral}
\chi^{JJ_z}_{\lambda_q\lambda_{\bar q}}&=&\sum_{m_\ell m_s}\langle L m_\ell; S m_s|J J_z\rangle\, Y_{Lm_\ell}(\hat{\mathbf{k}})\times\sum_{s\bar s}\mathcal{M}_{\lambda_q s}\,\mathcal{M}^{\bar q}_{\lambda_{\bar q}\bar s}\,\langle \tfrac12 s; \tfrac12 \bar s|S m_s\rangle .
\end{eqnarray}
Since $\mathcal{M}$ is unitary, Eq. (\ref{chigeneral}) is normalised point by point in $(x,\mathbf{k}_\perp)$. Setting $L=0$ returns to Eqs. (\ref{fockstate}) and (\ref{vector}) exactly, which is what fixes the phase conventions before the $L=1$ case is used. For the two $P$-wave states, whose covariant vertices were given in Ref. \cite{Cheng:2003sm}, we then obtain
\begin{eqnarray}\label{hcwave}
\psi^{1J_z}_{h_c}&=&\phi^{1P}_{m=J_z}(x,\mathbf{k}_\perp)\,\chi^{00}_{\lambda_q\lambda_{\bar q}}(x,\mathbf{k}_\perp),
\end{eqnarray}
for $h_c$ ($L=1$, $S=0$, $J=1$), where the spin part is the singlet and the angular momentum resides entirely in the orbital motion, and
\begin{eqnarray}\label{chicwave}
\psi^{00}_{\chi_{c0}}&=&\sum_{m=-1}^{1}\langle 1\,m;1\,{-m}|0\,0\rangle\,\phi^{1P}_{m}(x,\mathbf{k}_\perp)\,\chi^{1,-m}_{\lambda_q\lambda_{\bar q}}(x,\mathbf{k}_\perp),
\end{eqnarray}
for $\chi_{c0}$ ($L=1$, $S=1$, $J=0$), with $\langle 1\,m;1\,{-m}|0\,0\rangle=(-1)^{1-m}/\sqrt3$. The $\chi_{c0}$ case is more subtle among the two, because orbital and spin degrees of freedom are entangled. The state $\psi_{\chi_{c0}}$ does not factorise into one orbital function times a fixed spin matrix and all three substates $m=0,\pm1$ of $\phi^{1P}$ contribute coherently. Both wave functions come out normalised without further adjustment. The radial wave function of $\phi_{nS(nP)}$ in Eq. (\ref{wave}) for $1S$, $2S$ and $1P$ states have the form as \cite{Hwang:2008qi, Ke:2010vn,Cheng:2003sm}
\begin{equation}
    \phi^{1S}=4\left( {\frac{\pi}{\beta^2}} \right)^{\frac{3}{4}}\sqrt{\frac{dk_z}{dx}}exp\left(-\frac{k_z^2+\mathbf{k}_\perp^2}{2\beta^2}\right), 
\end{equation}
\begin{equation}
    \phi^{2S}=4\left( {\frac{\pi}{\beta^2}} \right)^{\frac{3}{4}}\sqrt{\frac{dk_z}{dx}}exp\left(-\frac{2^\delta}{2}\frac{k_z^2+\mathbf{k}_\perp^2}{\beta^2}\right)\left(a^\prime_2-b_2^\prime\frac{k_z^2+\mathbf{k}_\perp^2}{\beta^2}\right), 
\end{equation}
\begin{equation}
    \phi^{1P}=4\sqrt{2}\left( {\frac{\pi}{\beta^2}} \right)^{\frac{3}{4}}\sqrt{\frac{dk_z}{dx}}\frac{k_m}{\beta}exp\left(-\frac{k_z^2+\mathbf{k}_\perp^2}{2\beta^2}\right), 
\end{equation}
where $k_m$ ($k_{\pm1}=\mp(k_{\perp1}\pm ik_{\perp2})/\sqrt{2}$ and $ k_{0}=k_z$), $m$ is the magnetic quantum number, which is obtained from spherical harmonics. In the above equation, the value of  $a^\prime_2$ is $1.88684$ and $ b^\prime_2$ is $1.54943$. The exponent parameter is $\delta=0.549453$. It is not a free parameter, being fixed together with $a_2^\prime$ and $b_2^\prime$ by requiring $\phi^{2S}$ to be orthogonal to $\phi^{1S}$ at the same $\beta$ and correctly normalised, which reproduces the two quoted constants to six digits. The magnetic quantum number $m$ is not a free choice. It is fixed by the angular-momentum coupling in Eqs. (\ref{hcwave}) and (\ref{chicwave}), and summed over where that coupling requires it. $\beta$ is the harmonic scale parameter, which is obtained by fitting with meson mass through the variational principle. The normalization condition for the radial wave function  $\phi^{1S}, \phi^{2S}, \phi^{1P}$ is 
\begin{equation}
    \int_{0}^{1} dx\int d^2\mathbf{k}_\perp\,{|\phi_{nS(nP)}(x,  \mathbf{k}_\perp)|^2} = 2(2\pi)^3 .
\end{equation}
We can express three momentum $\mathbf{k}=(k_z,  \mathbf{k}_\perp)$ as $\mathbf{k}=(x, \mathbf{k}_\perp)$ through the relation 
\begin{equation}
    k_z=\left(x-\frac{1}{2}\right)M_o+\frac{m_{\bar{q}}^2-m_q^2}{2M_o} .
\end{equation}
The Jacobian factor of the variable transformation $(k_z,\mathbf{k}_\perp)\rightarrow(x,  \mathbf{k}_\perp)$ is given as

\begin{equation}
    \frac{dk_z}{dx} = \frac{M_o}{4x(1-x)} \left[ 1 - \frac{\left(m_q^2 - m_{\bar{q}}^2\right)^2}{M_o^4} \right].
\end{equation}
\subsection{Decay constant}
The pseudo-scalar and vector meson decay constant is defined through the matrix
elements of axial-vector and vector currents between the meson state and vacuum with mass $M$ and four-vector momenta $P^\mu$ as \cite{Choi:2007yu,Ahmady:2018muv}
\begin{eqnarray}
      \langle0|{\bar q_1\gamma^\mu\gamma_5q_2}|\mathcal{P}(P)\rangle=i{f_\mathcal{P}}P^\mu, 
\end{eqnarray}
\begin{eqnarray}
    \langle0|{\bar q_1\gamma^\mu q_2}|\mathcal{V}(P,\lambda)\rangle =i{f_\mathcal{V}}M_\mathcal{V}\,\epsilon ^\mu (\lambda). 
\end{eqnarray}
Here, $f_\mathcal{P(V)}$ is the decay constant of pseudo-scalar (vector) mesons. $q_{1(2)}$ are the quark flavors inside the mesons.
The decay constant for the pseudo-scalar meson can be expressed explicitly as  
\begin{eqnarray}
    f_{\mathcal{P}}=\sqrt{6}\int dx\int \frac{d^2\mathbf{k}_\perp}{(2\pi)^3} \frac{\phi_{nS(nP)}{(x, \mathbf{k}_\perp)}}{\sqrt{A^2+\mathbf{k}_\perp^2}}\mathcal{U_\mathcal{P}},\label{decayP}
\end{eqnarray}
where $A=x\,m_{\bar q} + (1-x)\,m_q$ and $\mathcal{U_{\mathcal{P}}}=A$.
\par For the vector meson, the `+' current component with longitudinal polarization $\epsilon(0)$ is completely identical as the `$\perp$' current component having transverse polarization $\epsilon(\pm)$ \cite{Choi:2013mda}.  The quantitative representation of vector decay constant is
\begin{eqnarray}
    f^{(+, \perp)}_{\mathcal{V}}=\sqrt{6}\int dx\int \frac{d^2\mathbf{k}_\perp}{(2\pi)^3} \frac{\phi_{nS(nP)}{(x,\mathbf{k}_\perp)}}{\sqrt{A^2+\mathbf{k}_\perp^2}}\mathcal{U^{(+, \perp)}_\mathcal{V}},\label{decayV}
\end{eqnarray}
where 
\begin{eqnarray}
\mathcal{U^{+}_{\mathcal{V}}}=A+\frac{2\mathbf{k}_\perp^2}{M_o+m_q+m_{\bar q}}, 
 \mathcal{U^{\perp}_{\mathcal{V}}}=\frac{1}{M_o}\left[ \frac{\mathbf{k}_\perp^2 + A^2}{2x(1-x)}-\mathbf{k}_\perp^2 + \frac{(m_q+m_{\bar q})}{M_o+m_q+m_{\bar q}}\mathbf{k}_\perp^2\right].
\end{eqnarray}
In the LFQM, we have successfully calculated that $f_\mathcal{V}=f^{+}_{\mathcal{V}}=f^{\perp}_{\mathcal{V}}$ for all radially excited states of mesons.

\par Equations (\ref{decayP}) and (\ref{decayV}) apply to the $S$-wave pseudo-scalar and vector states. The $P$-wave states $\chi_{c0}$ and $h_c$ also appear in Table \ref{decayconstant} permitted by discrete symmetries. For a $q\bar q$ state, built from equal-mass quarks, the parity and charge conjugation eigenvalues are \cite{ParticleDataGroup:2022pth}
\begin{eqnarray}\label{cparity}
\mathcal{P}=(-1)^{L+1},&&\qquad \mathcal{C}=(-1)^{L+S},
\end{eqnarray}
so that the four states entering this work carry
\begin{eqnarray}
\eta_c\ (^1S_0):&&\ L=0,\,S=0\ \Rightarrow\ J^{PC}=0^{-+},\nonumber\\
J/\psi\ (^3S_1):&&\ L=0,\,S=1\ \Rightarrow\ J^{PC}=1^{--},\nonumber\\
\chi_{c0}\ (^3P_0):&&\ L=1,\,S=1\ \Rightarrow\ J^{PC}=0^{++},\nonumber\\
h_c\ (^1P_1):&&\ L=1,\,S=0\ \Rightarrow\ J^{PC}=1^{+-}.
\end{eqnarray}
The interpolating currents have the definite charge conjugation \cite{Li:2026pwave}
\begin{eqnarray}\label{ccurrents}
\mathcal{C}\big(\bar q\gamma^{\mu}q\big)=-1,&&\qquad \mathcal{C}\big(\bar q\gamma^{\mu}\gamma_5 q\big)=+1,\nonumber\\
\mathcal{C}\big(\bar q\sigma^{\mu\nu}q\big)=-1,&&\qquad \mathcal{C}\big(\bar q q\big)=+1.
\end{eqnarray}
A local matrix element $\langle0|J|M\rangle$ can be non-zero only when $\mathcal{C}(J)=\mathcal{C}(M)$. The rule is satisfied in the two familiar $S$-wave cases, $\mathcal{C}(\bar q\gamma^{\mu}\gamma_5q)=\mathcal{C}(\eta_c)=+1$ and $\mathcal{C}(\bar q\gamma^{\mu}q)=\mathcal{C}(J/\psi)=-1$. For the $P$-wave states it fails in both the channels, $\mathcal{C}(\bar q\gamma^{\mu}q)=-1\neq\mathcal{C}(\chi_{c0})=+1$ and $\mathcal{C}(\bar q\gamma^{\mu}\gamma_5q)=+1\neq\mathcal{C}(h_c)=-1$, and the two local matrix elements vanish identically
\begin{eqnarray}\label{cvanish}
\langle 0|\bar q\gamma^{\mu} q|\chi_{c0}(P)\rangle = 0, &&
\langle 0|\bar q\gamma^{\mu}\gamma_5 q|h_c(P,\lambda)\rangle = 0.
\end{eqnarray}
The associated constants are then not decay constants in the usual sense. What survives is the normalisation of the leading-twist light-cone distribution amplitude, obtained from the light-like separated correlators \cite{Li:2026pwave}
\begin{eqnarray}\label{fscalar}
f_{\chi_{c0}}\,\phi_{\chi_{c0}}(x)&=&\sqrt{\frac{2N_c}{x(1-x)}}\int\frac{d^2\mathbf{k}_\perp}{(2\pi)^3}\,
\psi^{\uparrow\downarrow+\downarrow\uparrow}_{\chi_{c0}}(x,\mathbf{k}_\perp),\\
\label{faxial}
f_{h_c}\,\phi_{h_c}(x)&=&\sqrt{\frac{2N_c}{x(1-x)}}\int\frac{d^2\mathbf{k}_\perp}{(2\pi)^3}\,
\psi^{(J_z=0)\,\uparrow\downarrow-\downarrow\uparrow}_{h_c}(x,\mathbf{k}_\perp).
\end{eqnarray}
Equation (\ref{cvanish}) makes the zeroth moment of these amplitudes vanish, so they have to be normalised by the first moment instead, $\int_0^1 dx\,(2x-1)\phi(x)=1$. For $h_c$ there is one current that does match, the tensor current $\bar q\sigma^{\mu\nu}q$ with $\mathcal{C}=-1$, and it defines a genuine decay constant
\begin{eqnarray}\label{ftensor}
\langle 0|\bar q\sigma^{\mu\nu}q|h_c(P,\lambda)\rangle&=&i f^{\perp}_{h_c}\big(\epsilon^{\mu}(\lambda)P^{\nu}-\epsilon^{\nu}(\lambda)P^{\mu}\big),\nonumber\\
f^{\perp}_{h_c}\,\phi^{\perp}_{h_c}(x)&=&\sqrt{\frac{2N_c}{x(1-x)}}\int\frac{d^2\mathbf{k}_\perp}{(2\pi)^3}\,
\psi^{(J_z=+1)\,\uparrow\uparrow}_{h_c}(x,\mathbf{k}_\perp),
\end{eqnarray}
normalised by $\int_0^1 dx\,\phi^{\perp}_{h_c}(x)=1$. Here $N_c=3$, the spin labels denote the corresponding combinations of the helicity components of Eq. (\ref{wave}), and $\psi=\sqrt{x(1-x)}\,\phi_{nS(nP)}\,\chi^{JJ_z}_{\lambda_q\lambda_{\bar q}}$. The wave functions of Eqs. (\ref{hcwave}) and (\ref{chicwave}) satisfy Eq. (\ref{cvanish}) identically, the contributions of the individual helicity components cancelling among themselves, while the same projections applied to the $S$-wave states return Eqs. (\ref{decayP}) and (\ref{decayV}).

\begin{table}
    \centering
    \begin{tabular}{|c|c|c|c|c|}
    \hline
         & $\chi_{c0}(1P)$ & $h_c(1P)$ &   $\eta_c(2S)$ & $\psi(2S)$ \\\hline
        Our result & 85. 1$^{\,a}$ & 195. 2$^{\,b}$ &203. 5 &266. 2 \\\hline
        Exp. \cite{ParticleDataGroup:2020ssz} & - & - & - & 294\\\hline
        LFQM \cite{Acharyya:2024enp} & 421 & 373 & 318 & 420\\\hline
         BS \cite{Zhou:2020bnm}& 239 & - & - & -\\\hline
         Sum rule \cite{Wang:2012gj}& - & 490 & - & -\\\hline
         BLFQ \cite{Li:2017mlw}& - & - & 298 &312 \\\hline
         LFD \cite{Li:2021cwv} & - & - & - &288 \\\hline
    \end{tabular} 
    
    \caption{Decay constant (in units of MeV) of excited state meson $\chi_{c0}(1P), h_c(1P), \eta_c(2S)$ and $\psi(2S)$ compared with experimental data \cite{ParticleDataGroup:2020ssz} and other model calculations of BS \cite{Zhou:2020bnm}, Sum rule \cite{Wang:2012gj}, BLFQ \cite{Li:2017mlw}, LFD \cite{Li:2021cwv} and LFQM \cite{Acharyya:2024enp}. Entries marked $^a$ are normalisations of the leading-twist light-cone distribution amplitude, fixed by the first moment, Eqs. (\ref{fscalar})--(\ref{faxial}). The entry marked $^b$ is the tensor decay constant $f^{\perp}_{h_c}$ of Eq. (\ref{ftensor}). For $\chi_{c0}(1P)$ and $h_c(1P)$ no decay constant of the conventional type exists, since charge conjugation forbids the corresponding local matrix elements.}
    \label{decayconstant}
\end{table}

\subsection{Transition Form Factor}
\begin{figure}
    \centering
    \includegraphics[width=0.8\linewidth]{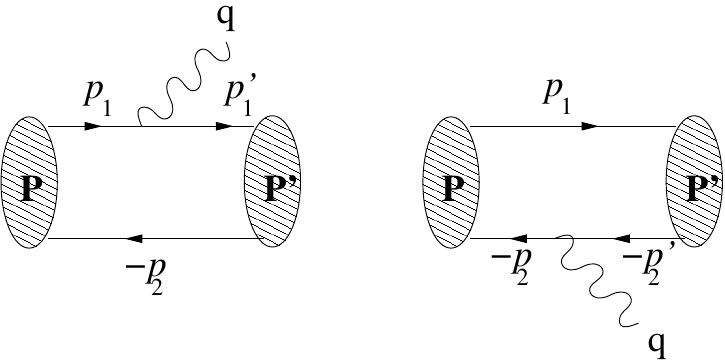}
    \caption{Lowest order Feynman diagram of Vector to pseudo-scalar charmonia.}
    \label{trans}
\end{figure}
In the LFQM, the E1 transition between a $P$-wave and a $S$-wave charmonium, $\mathcal{S}\leftrightarrow\mathcal{V}\gamma^*$, proceeds through the diagram of Fig. \ref{trans}. Throughout, $\mathcal{S}$ denotes the spin-zero member of the pair, $\chi_{c0}$ for the $0^{++}\leftrightarrow1^{--}$ family and $\eta_c$ for the $0^{-+}\leftrightarrow1^{+-}$ family, and $\mathcal{V}$ the spin-one member, $J/\psi$, $\psi(2S)$ or $h_c$. The E1 matrix element does not have the M1 form $\epsilon^{\mu\nu\rho\sigma}\epsilon_\nu q_\rho P_\sigma$. That structure belongs to a magnetic dipole and cannot describe a transition in which the orbital angular momentum changes by one unit at fixed quark spin. The decomposition into Lorentz structures appropriate to E1 reads \cite{Wang:2023nhb,Dudek:2006ej}
\begin{eqnarray}\label{e1decomp}
&&\langle \mathcal{V}(P',\lambda')|J^\mu_{em}(0)|\mathcal{S}(P)\rangle=E_1(Q^2)\Bigg[\epsilon^{*\mu}_{\lambda'}(P')-\frac{\epsilon^{*}_{\lambda'}\cdot P}{\Omega(Q^2)}\Big(P'^{\mu}(P\cdot P')-M_\mathcal{V}^2P^\mu\Big)\Bigg]\nonumber\\
&&\qquad+\frac{M_\mathcal{V}C_1(Q^2)}{i Q\,\Omega(Q^2)}\big(\epsilon^{*}_{\lambda'}\cdot P\big)\times\Big[(P\cdot P')(P+P')^\mu-M_\mathcal{S}^2P'^\mu-M_\mathcal{V}^2P^\mu\Big],
\end{eqnarray}
where $q=P'-P$ is the four-momentum of the virtual photon, $Q^2=-q^2$, $\Omega(Q^2)=(P\cdot P')^2-M_\mathcal{S}^2M_\mathcal{V}^2$, and $\epsilon^\mu_{\lambda'}(P')$ is the polarisation vector of the spin-one state. $E_1(Q^2)$ and $C_1(Q^2)$ are the transverse (electric dipole) and longitudinal (Coulomb) form factors, and $E_1$ has the dimension of a mass. The same decomposition holds for the $0^{-+}\leftrightarrow1^{+-}$ pair.

The calculation is performed in the Drell-Yan frame $q^+=0$ with $\mathbf{P}_\perp=0$ and $P^+=1$, such that
\begin{eqnarray}
P&=&(1,\,M_\mathcal{S}^2,\,\mathbf{0}_\perp),\qquad P'=(1,\,M_\mathcal{V}^2+Q^2,\,\mathbf{q}_\perp),\nonumber\\
Q^2&=&-q^2=\mathbf{q}_\perp^2\;\ge 0,
\end{eqnarray}
i.e. the space-like region, which for the real-photon point is reached as $Q^2\rightarrow0$. Only the `good' component $\mu=+$ of the current is used, making the result free of zero-mode contributions. Writing $A_{\lambda'}=\langle \mathcal{V}(P',\lambda')|J^+_{em}|\mathcal{S}(P)\rangle$ and $D=M_\mathcal{S}^2-M_\mathcal{V}^2+Q^2$, an explicit evaluation of Eq. (\ref{e1decomp}) for the longitudinal ($\lambda'=0$) and transverse ($\lambda'=+1$) polarisations gives
\begin{eqnarray}
A_0&=&E_1\left[\frac{1}{M_\mathcal{V}}-\frac{D^2}{4M_\mathcal{V}\Omega}\right]+\frac{C_1QD}{2i\Omega},\nonumber\\
A_{+1}&=&\frac{E_1QD}{2\sqrt2\,\Omega}-\frac{M_\mathcal{V}C_1Q^2}{i\sqrt2\,\Omega}.
\end{eqnarray}
Neither amplitude on its own isolates $E_1$, since both carry a $C_1$ admixture. However, the combination
\begin{eqnarray}\label{e1combo}
E_1(Q^2)&=&M_\mathcal{V}A_0+\frac{M_\mathcal{S}^2-M_\mathcal{V}^2+Q^2}{\sqrt2\,Q}\,A_{+1}
\end{eqnarray}
isolates it as the $C_1$ terms cancel identically for any $Q^2$ and any pair of masses, leaving $E_1$ with unit coefficient. Equation (\ref{e1combo}) plays the role for E1 that the several current components play for M1 making an extraction from the plus current possible.

\par In the valence sector, the plus current is diagonal in the quark helicity and brings no spin-flip Dirac factor, so each $A_{\lambda'}$ reduces to a plain overlap of light-front wave functions. Momentum is important since it determines where the transverse shift acts. With the initial spin-zero state at $\mathbf{P}_\perp=0$ the struck quark carries $\mathbf{k}_\perp$ and absorbs $\mathbf{q}_\perp$, while the final state has $\mathbf{P}'_\perp=\mathbf{q}_\perp$, so the relative transverse momentum in the final state is $\mathbf{k}^\prime_\perp=\mathbf{k}_\perp+(1-x)\mathbf{q}_\perp$. The shift therefore belongs on the final-state wave function, giving
\begin{eqnarray}\label{overlap}
A_{\lambda'}&=&2\sum_{\lambda_q\lambda_{\bar q}}\int_0^1 dx\int\frac{d^2\mathbf{k}_\perp}{(2\pi)^3}\psi^{(\lambda')*}_{\mathcal{V}}\big(x,\mathbf{k}_\perp+(1-x)\mathbf{q}_\perp\big)\,\psi_{\mathcal{S}}(x,\mathbf{k}_\perp),
\end{eqnarray}
where the overall factor $4/2$ arises as follows. There is a factor $2$ from the current normalisation $\langle P'|J^+|P\rangle\propto2P^+$ with $P^+=1$, a further factor $2$ from the two diagrams in which the photon attaches to the quark and to the antiquark (equal for equal quark masses and opposite $C$-parity of $\mathcal{S}$ and $\mathcal{V}$), and a $1/2$ from combining the measure of Eq. (\ref{fockstate0}) with $\psi=\sqrt{x(1-x)}\,\phi_{nS(nP)}\,\chi^{JJ_z}$. Consistently with Eq. (\ref{e1decomp}), $J^{\mu}_{em}$ denotes the charge-stripped current, the quark charge $\mathcal{Q}_c=2/3$ being factored out and reinstated in Eq. (\ref{e1width}). Equations (\ref{e1combo}) and (\ref{overlap}), together with the wave functions of Eqs. (\ref{hcwave}) and (\ref{chicwave}), constitute the complete E1 formalism used in this work.

\par The real-photon point needs a limiting procedure. As $Q\rightarrow0$ the transverse amplitude vanishes linearly, $A_{+1}=\mathcal{O}(Q)$, while the coefficient multiplying it in Eq. (\ref{e1combo}) diverges as $1/Q$, so $E_1(0)$ cannot be evaluated directly at $Q=0$. Since $E_1$ is a smooth even function of $Q$, we obtain it by fitting a quadratic in $Q^2$ over $Q\in[0.10,0.30]$ GeV and extrapolating. The coupling constant for a real photon is $\mathcal{G}=E_1(0)$, and the E1 width is \cite{Wang:2023nhb}
\begin{eqnarray}\label{e1width}
\Gamma_{(i\rightarrow f\gamma)}&=&\frac{\mathcal{Q}_c^2\,\alpha}{2j_i+1}\,\frac{M_i^2-M_f^2}{2M_i^3}\,\big|E_1(0)\big|^2,
\end{eqnarray}
where $\alpha$ is the fine structure constant, $\mathcal{Q}_c=2/3$ is the charm quark charge number, and $j_i$, $M_i$ and $M_f$ are the spin of the decaying state and the initial and final masses. Equation (\ref{e1width}) is not of the $\mathcal{G}^2K_\gamma^3$ form appropriate to an M1 transition. Since $E_1$ carries the dimension of a mass, the phase-space factor for an electric dipole is linear, rather than cubic, in the photon energy.

\subsection{Branching ratio}
A particle can decay in multiple modes, each of which has its own decay rate. The overall decay rate is the sum of the decay rates of all the modes. The total decay modes can be expressed as  
\begin{eqnarray}
    \Gamma_{total}= \sum_{i=1}^{n} \Gamma_i . 
\end{eqnarray}
Lifetime is associated with the total decay modes as  $\tau=1/\Gamma_{total}$. 
It is also intriguing to determine the branching ratio or proportion, or the likelihood of decay by a specific decay mode, if there are multiple decay modes available. The branching fraction of $i_{th}$ mode can be written as 
\begin{eqnarray}
    B_i=\Gamma_i/\Gamma_{total}.
\end{eqnarray}
\section{Result and Discussion}\label{results}
\begin{figure}[ht]
		\centering
		\begin{minipage}[c]{1\textwidth}\begin{center}
				(a)\includegraphics[width=.45\textwidth]{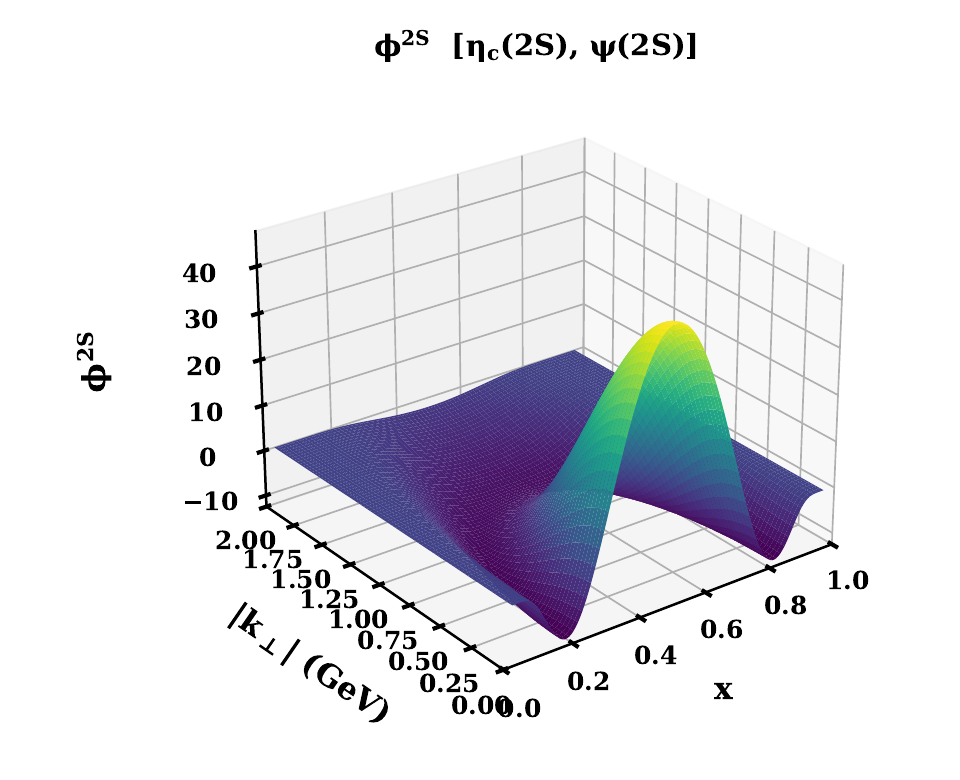}
				(b)\includegraphics[width=.45\textwidth]{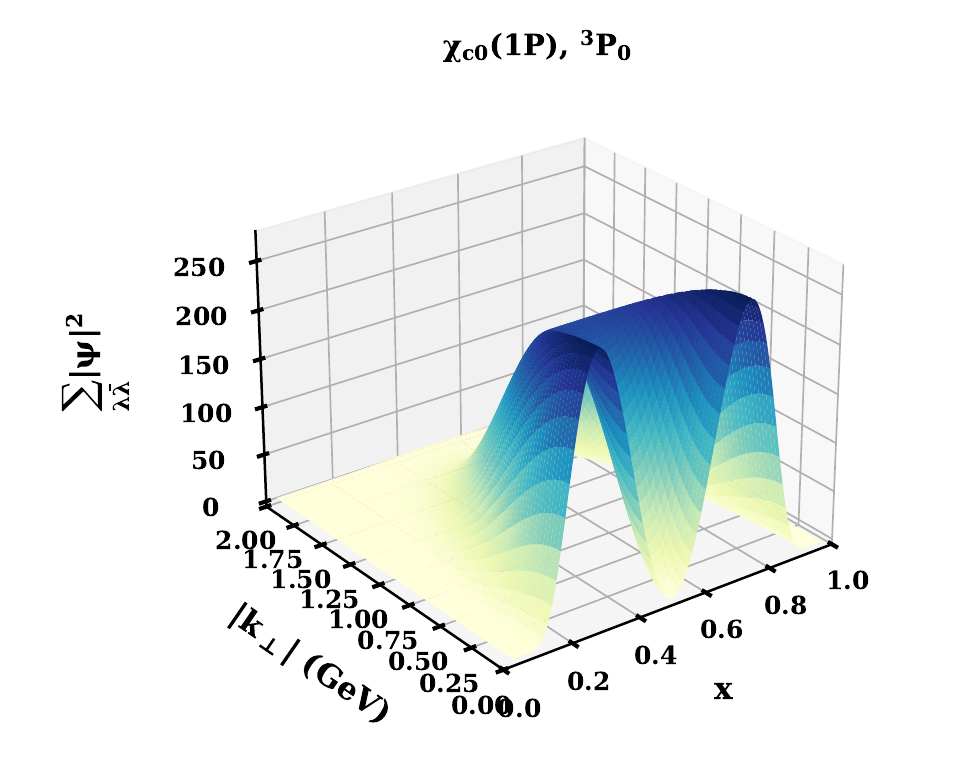}
			\end{center}
		\end{minipage}
		\begin{minipage}[c]{1\textwidth}\begin{center}
				(c)\includegraphics[width=.45\textwidth]{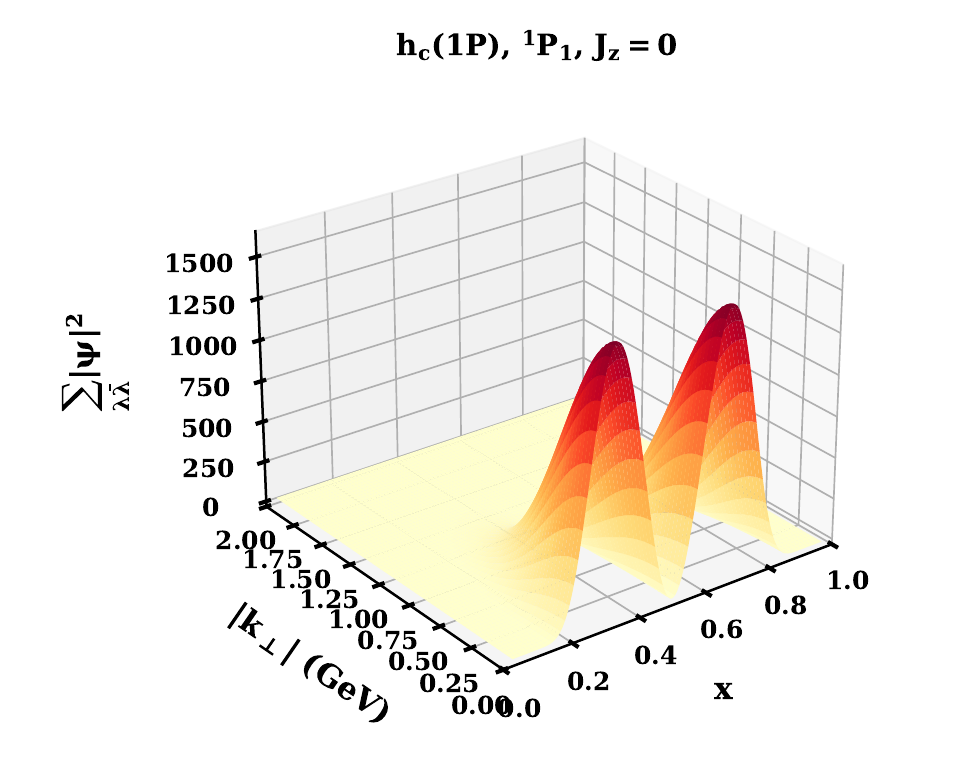}
                (d)\includegraphics[width=.45\textwidth]{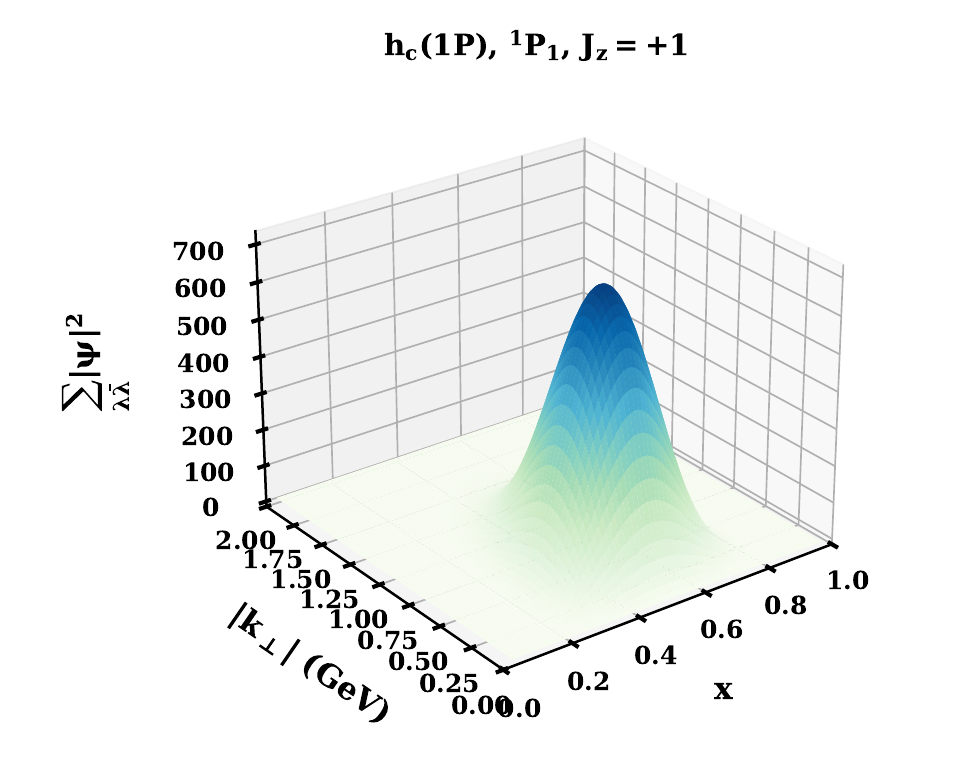}
			\end{center}
		\end{minipage}
		\caption{(Color online) Light-front wave functions with respect to the longitudinal momentum fraction $x$ and the transverse momentum $\bfk$. Panel (a) is the radial function $\phi^{2S}$, which is common to $\eta_c(2S)$ and $\psi(2S)$. Panels (b)--(d) are the helicity-summed densities $\sum_{\lambda_q\lambda_{\bar q}}|\psi|^2$ of the physical $P$-wave states, for $\chi_{c0}(1P)$ and for $h_c(1P)$ with $J_z=0$ and $J_z=+1$.}
		\label{21realtmds3}
	\end{figure}

For the numerical predictions of our calculations, we require only two input parameters: charm quark mass $m_c$ and harmonic scale parameter $\beta$. Both are fixed by the variational principle, by minimising $\langle H_0\rangle+\langle V_{\rm conf}\rangle+\langle V_{\rm Coul}\rangle+\langle V_{\rm hyp}\rangle$ and requiring the two measured $1S$ masses to be reproduced. The two conditions $M=M_{\eta_c(1S)}$ and $M=M_{J/\psi}$ determine $m_c=1.6012$ GeV and $\beta=0.7109$ GeV exactly, and the hyperfine splitting then comes out at its measured value of $113.00$ MeV. We use the same $\beta$ for every charmonium state, so that no parameter is tuned to the transitions themselves. Figure \ref{21realtmds3} shows the light-front wave functions against the longitudinal momentum fraction $x$ and the transverse momentum $\bfk$ of the active quark. Panel (a) is the radial function $\phi^{2S}(x,\bfk)$, with its node clearly visible. The same function serves both $\eta_c(2S)$ and $\psi(2S)$, because the two states differ only in their spin-orbit functions and those are normalised to unity at every $(x,\bfk)$. Panels (b)--(d) show the helicity-summed density $\sum_{\lambda_q\lambda_{\bar q}}|\psi|^2$ for the $P$-wave states. In panel (b) the coherent sum over $m$ in Eq. (\ref{chicwave}) leaves $\chi_{c0}(1P)$ with a density proportional to $k_z^2+\bfk^2$, so it vanishes only at the single point $x=1/2$, $\bfk=0$. The $h_c(1P)$ density, by contrast, depends on the polarisation, and panels (c) and (d) show $J_z=0$ and $J_z=+1$. The $J_z=0$ density carries a factor $k_z^2$ and is suppressed along the whole line $x=1/2$, while the $J_z=+1$ density carries $\bfk^2/2$ and is suppressed at small $\bfk$ instead. This difference between the two polarisations follows directly from the angular-momentum coupling of Sec. \ref{satya}, and an $S$-wave treatment has nothing corresponding to it.

\par In Table \ref{decayconstant} we present the decay constants. For the $2S$ states these are conventional decay constants, and our values $f_{\eta_c(2S)}=203.5$ MeV and $f_{\psi(2S)}=266.2$ MeV are comparable with the BLFQ \cite{Li:2017mlw} and LFD \cite{Li:2021cwv} results and lie somewhat below the measured $f_{\psi(2S)}=294$ MeV. Within this model $|f|$ for a $2S$ pseudo-scalar saturates near $213$ MeV for any $\beta$, so that values in the $300$ MeV range are not attainable with a single harmonic-oscillator $2S$ trial function. This is the known $nS$ decay-constant hierarchy problem of the LFQM \cite{Ke:2010vn}.

\par The $P$-wave states behave quite differently, and here we correct the previous version. The selection rule of Eqs. (\ref{cparity})--(\ref{ccurrents}) forbids the vector current from coupling to $\chi_{c0}$ and the axial-vector current from coupling to $h_c$, so neither state has a decay constant of the conventional type and Eq. (\ref{cvanish}) holds identically. The cancellation is not imposed by hand. Evaluating the zeroth moments with the wave functions of Eqs. (\ref{hcwave}) and (\ref{chicwave}), the contributions of the individual helicity components cancel among themselves and the two matrix elements vanish, as Eq. (\ref{cvanish}) requires. This provides an independent check on the angular-momentum coupling of the $P$-wave states, which would otherwise leave a finite residue. What can be quoted instead are the first-moment normalisations of the leading-twist distribution amplitudes, $f_{\chi_{c0}}=85.1$ MeV and $f_{h_c}=147.4$ MeV, and, for $h_c$ only, the genuine tensor decay constant $f^{\perp}_{h_c}=195.2$ MeV, the tensor current being the one with $\mathcal{C}=-1$. The values $426.4$ and $317.39$ MeV quoted previously came from applying the $S$-wave formulas (\ref{decayP}) and (\ref{decayV}) to $P$-wave states, and do not correspond to quantities that exist. Applied to states where the conventional definition does hold, the same machinery gives $f_{\eta_c(1S)}=364.3$ MeV and $f_{J/\psi}=416.8$ MeV, against the measured $\simeq335$ MeV and $407\pm5$ MeV. Neither entered the fit. 

\begin{figure}[ht]
		\centering
		\begin{minipage}[c]{1\textwidth}\begin{center}
				(a)\includegraphics[width=.45\textwidth]{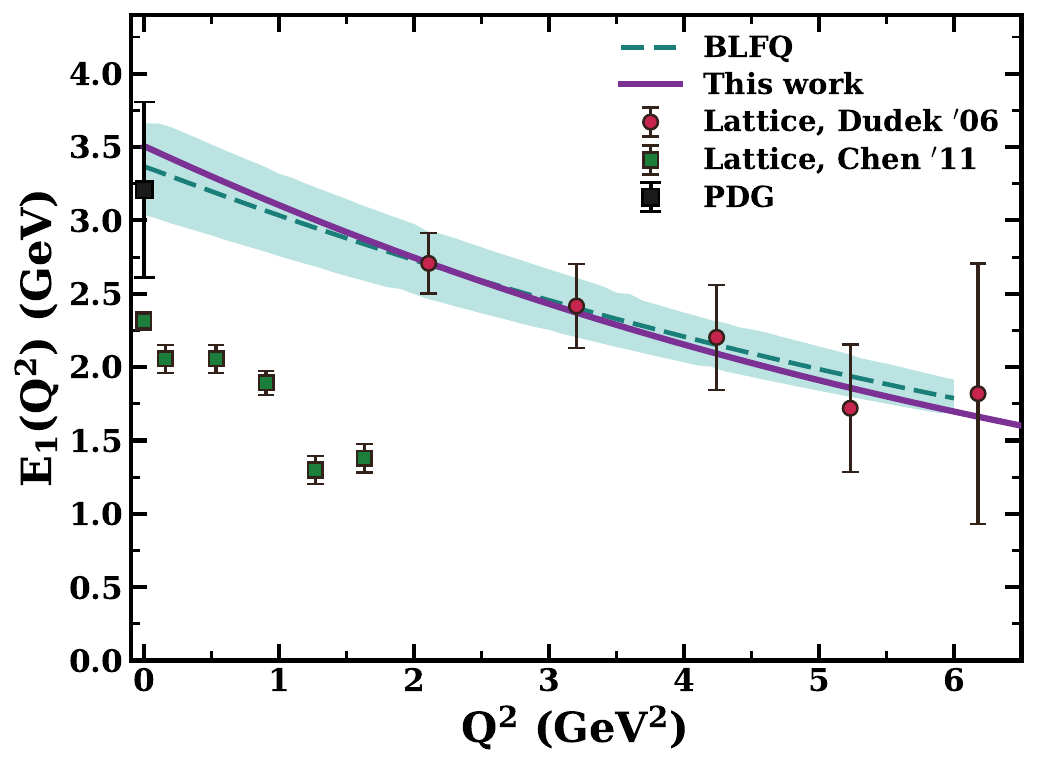}
				(b)\includegraphics[width=.45\textwidth]{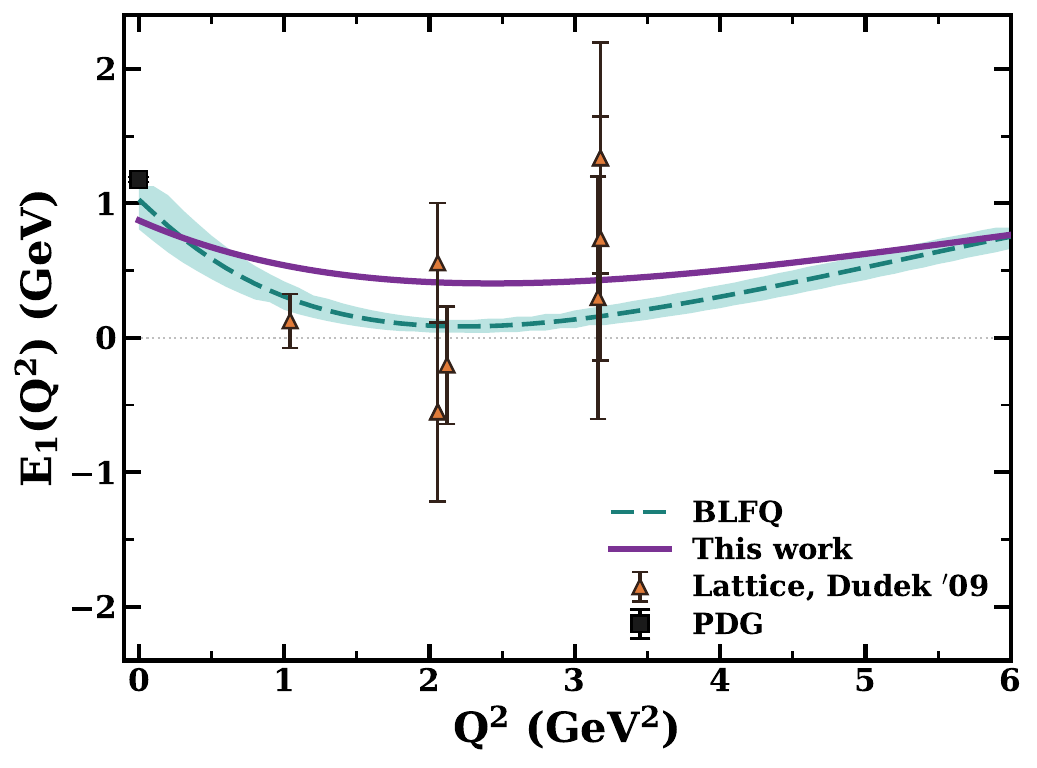}
			\end{center}
		\end{minipage}
		\begin{minipage}[c]{1\textwidth}\begin{center}
				(c)\includegraphics[width=.45\textwidth]{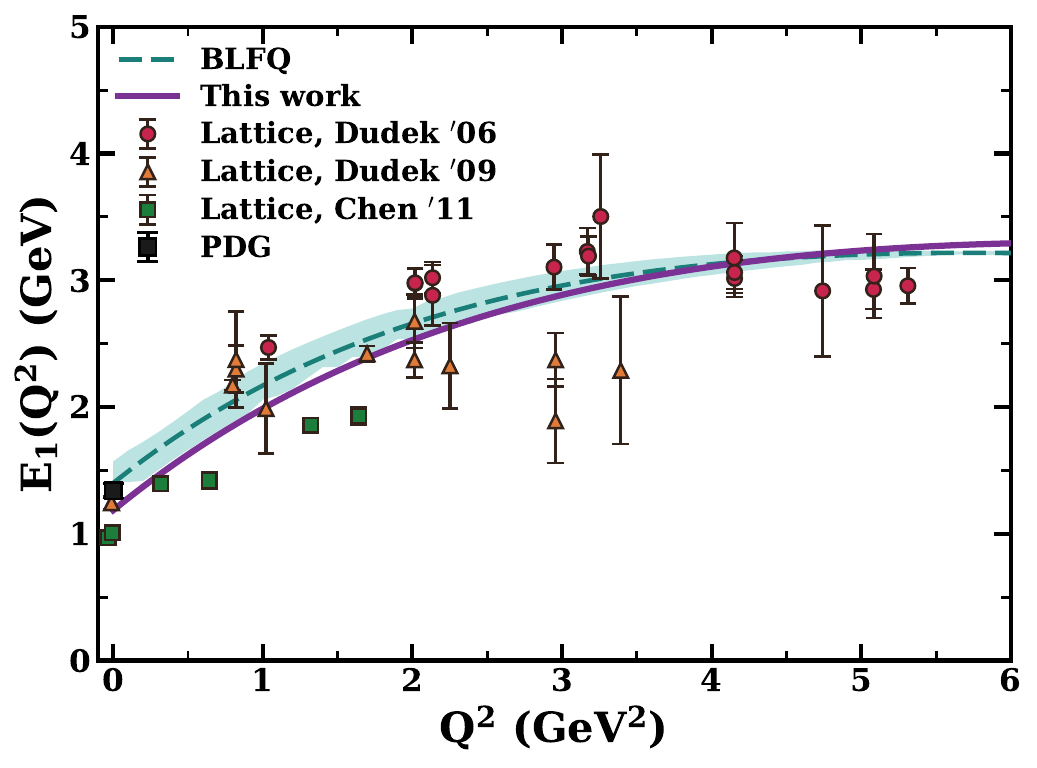}
				(d)\includegraphics[width=.45\textwidth]{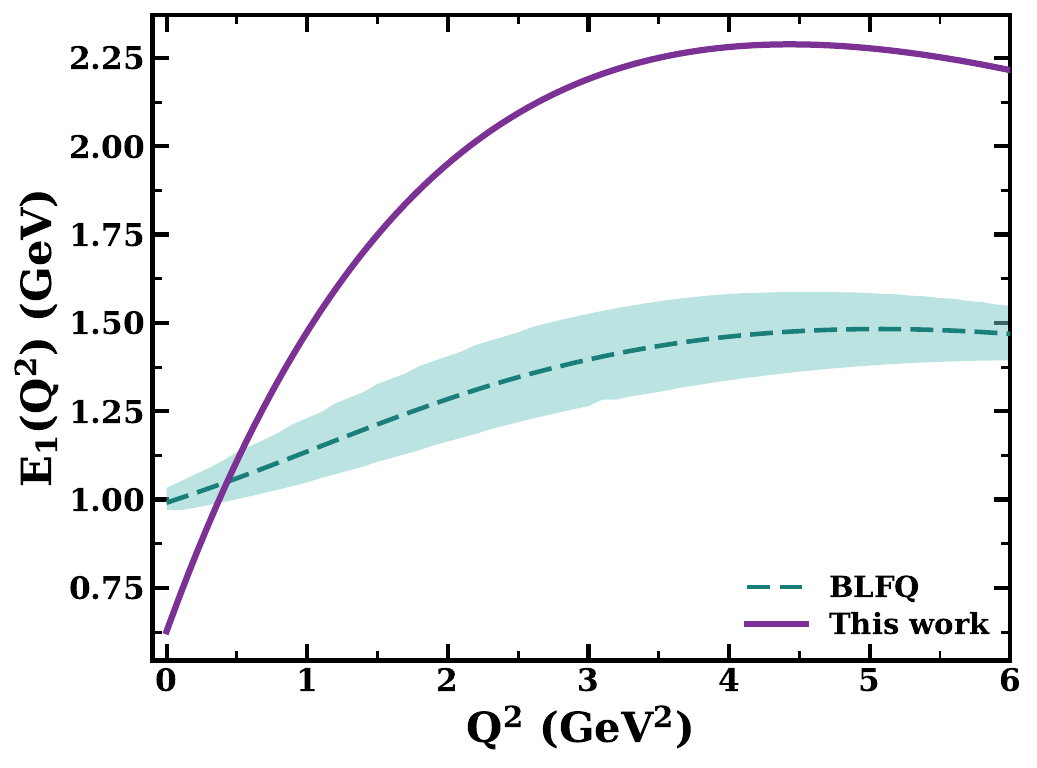}
			\end{center}
		\end{minipage}
		\caption{(Color online) The electric dipole transition form factors $E_1(Q^2)$ plotted with respect to $Q^2$ (GeV$^2$) for (a) the $h_c(1P)\rightarrow\eta_c(1S)+\gamma$ transition, compared with the lattice simulations of Refs. \cite{Dudek:2006ej,Chen:2011kpa} and the PDG \cite{ParticleDataGroup:2022pth} value at $Q^2=0$, (b) the $\psi(2S)\rightarrow \chi_{c0}(1P)+\gamma$ transition, compared with the lattice simulation of Ref. \cite{Dudek:2009kk} and the PDG \cite{ParticleDataGroup:2022pth} value at $Q^2=0$, (c) the $\chi_{c0}(1P)\rightarrow J/\psi(1S)+\gamma$ transition, compared with the lattice simulations of Refs. \cite{Dudek:2006ej,Dudek:2009kk,Chen:2011kpa} and the value at $Q^2=0$ obtained from the PDG width \cite{ParticleDataGroup:2022pth}, and (d) the $\eta_c(2S)\rightarrow h_c(1P)+\gamma$ transition, for which no lattice result is available. The solid curve is the present calculation. For BLFQ \cite{Wang:2023nhb} both the central value and the basis-sensitivity band are shown, in all panels.}
		\label{realtmds6}
	\end{figure}

In this work, we have mainly targeted the radiative transitions between the $P$-wave and $S$-wave charmonia. We have considered $4$ radiative transitions between pseudo-scalar and vector meson charmonia. In Fig. \ref{realtmds6}, we have plotted the $E_1(Q^2)$ TFFs of all four transitions with respect to $Q^2$ (in GeV$^2$), (a) $h_c(1P)\rightarrow\eta_c(1S)+\gamma$, (b) $\psi(2S)\rightarrow \chi_{c0}(1P)+\gamma$, (c) $\chi_{c0}(1P)\rightarrow J/\psi(1S)+\gamma$ and (d) $\eta_c(2S)\rightarrow h_c(1P)+\gamma$, since the TFF ${E}_1(Q^2)$ provides more resolution of the system. The solid curve in Fig. \ref{realtmds6} is the present calculation. Since $m_c$ and $\beta$ are fixed by two masses that are themselves known to better than a MeV, the parameters carry almost no freedom, and we therefore quote a single curve. For BLFQ we show the central value together with the basis-sensitivity band, taken from Fig. 5 of Ref. \cite{Wang:2023nhb}. The previous version displayed BLFQ as a band in one panel and as a line in the other, which made the two comparisons look different when they are not.

\par For the $h_c(1P)\rightarrow\eta_c(1S)+\gamma$ transition in Fig. \ref{realtmds6}(a), our $E_1(Q^2)$ falls monotonically from $3.51$ GeV at $Q^2=0$ to $1.70$ GeV at $Q^2=6$ GeV$^2$, to be compared with $3.37\rightarrow1.79$ GeV in BLFQ, and the value at the real-photon point agrees with the PDG \cite{ParticleDataGroup:2022pth} value of $3.21\pm0.60$ GeV. The lattice results of Dudek \textit{et al.} \cite{Dudek:2006ej}, which are available for $2\lesssim Q^2\lesssim6$ GeV$^2$, lie on our curve within their errors over the whole range. The twisted-mass results of Ref. \cite{Chen:2011kpa} at $Q^2<2$ GeV$^2$, on the other hand, lie about $35$--$55\%$ below our curve, the BLFQ band and the PDG point. The two lattice calculations are therefore not consistent with each other in this channel, which is also visible in their decay widths in Table \ref{tab:my_label22}, $601$ keV \cite{Dudek:2006ej} against $234$ keV \cite{Chen:2011kpa}. Our result follows the points of Ref. \cite{Dudek:2006ej} and the PDG value. A third lattice result, the $N_f=2$ twisted-mass calculation of Ref. \cite{Becirevic:2012dc} extrapolated to the continuum limit, gives $720$ keV. Since Refs. \cite{Chen:2011kpa} and \cite{Becirevic:2012dc} both use $N_f=2$ twisted-mass configurations, the factor of three between them cannot come from quenching alone, and must lie in the discretisation and in the way $E_1(0)$ is extracted. Two of the three lattice results, Refs. \cite{Dudek:2006ej,Becirevic:2012dc}, are on the side of the light-front approaches, our $534.5$ keV and $489$ keV of BLFQ, and of the NR ($498$ keV) \cite{Barnes:2005pb} and RQM ($560$ keV) models, while Ref. \cite{Chen:2011kpa} is closer to the GI, LP, SP ($373$ keV) \cite{PhysRevD.95.034026}, SNR$_1$ and CCQM \cite{Issadykov:2025} values of $304$--$373$ keV. The model predictions themselves range from $304$ to $764$ keV and cannot settle the question. Experimentally, the branching ratio is already precise, $\mathcal{B}(h_c\rightarrow\gamma\eta_c)=(57.7^{+3.6}_{-3.5})\%$, but the total width $\Gamma(h_c)=0.78^{+0.27}_{-0.24}\pm0.12$ MeV \cite{BESIII:2022tfo} carries an error of about $35\%$. A more precise $\Gamma(h_c)$, as expected from the resonance scan in $\bar pp\rightarrow h_c$ at PANDA \cite{PANDA:2009yku} or from the large charmonium samples at a future super tau-charm facility \cite{Achasov:2023gey}, would discriminate between the lattice results of Refs. \cite{Chen:2011kpa} and \cite{Dudek:2006ej,Becirevic:2012dc}.

\par For the $\psi(2S)\rightarrow \chi_{c0}(1P)+\gamma$ transition in Fig. \ref{realtmds6}(b), the form factor first falls, passes through a minimum near $Q^2\simeq2.5$ GeV$^2$ and then rises again, which is the same shape as the BLFQ curve. The minimum reflects the node of the $2S$ wave function in the $2S$--$1P$ dipole overlap. Our minimum stays at about $0.4$ GeV, whereas the BLFQ curve almost touches zero. The lattice points of Ref. \cite{Dudek:2009kk} in this region scatter around zero with errors of $0.4$--$0.9$ GeV, and cannot yet distinguish between the two. At $Q^2=0$ we obtain $0.87$ GeV against $1.18$ GeV from the PDG.

\par Fig. \ref{realtmds6}(c) shows the $\chi_{c0}(1P)\rightarrow J/\psi(1S)+\gamma$ transition, for which lattice results are available from three groups \cite{Dudek:2006ej,Dudek:2009kk,Chen:2011kpa}. In contrast to the two transitions above, the form factor here increases with $Q^2$, from $1.19$ GeV at $Q^2=0$ to $3.29$ GeV at $Q^2=6$ GeV$^2$, against $1.40\rightarrow3.22$ GeV in BLFQ. The three lattice data sets again differ from one another. The points of Ref. \cite{Dudek:2006ej} lie highest, those of Ref. \cite{Chen:2011kpa} lowest and those of Ref. \cite{Dudek:2009kk} in between. Our curve lies within this spread. It is above the points of Ref. \cite{Chen:2011kpa} and below those of Ref. \cite{Dudek:2006ej} for $Q^2\lesssim3$ GeV$^2$, and agrees with the latter for $Q^2\gtrsim4$ GeV$^2$. At the real-photon point our value of $1.19$ GeV is close to the lattice value $1.24\pm0.05$ GeV of Ref. \cite{Dudek:2009kk} and about $10\%$ below the value $1.34\pm0.06$ GeV obtained from the PDG width.

\par For the $\eta_c(2S)\rightarrow h_c(1P)+\gamma$ transition in Fig. \ref{realtmds6}(d), no lattice result is available so far, and we compare only with BLFQ. Our form factor starts at $0.63$ GeV at $Q^2=0$, rises quickly up to about $2.3$ GeV near $Q^2\simeq4.5$ GeV$^2$ and then flattens, while BLFQ starts at $0.99$ GeV and rises slowly to about $1.5$ GeV. As in Fig. \ref{realtmds6}(b), the small value at $Q^2=0$ comes from the node of the $2S$ wave function, which cancels a large part of the dipole overlap, and it is the reason for the small decay width discussed below. This is the channel in which the two light-front approaches differ the most, and a lattice calculation of it would be very useful. The difference comes from the $2S$ wave function. In our model its node is fixed by orthogonality to the $1S$ state at a common $\beta$, while in BLFQ it is fixed dynamically by confinement and one-gluon exchange \cite{Li:2017mlw}. Since the $2S$--$1P$ dipole overlap nearly cancels between the two lobes of the $2S$ function, this small difference is strongly enhanced, and the width ratio $26.4/10.7\approx2.5$ is just $(0.99/0.63)^2$. The measured $\psi(2S)\rightarrow\chi_{c0}(1P)+\gamma$ channel, which has the same overlap and where our width is about half of the experimental one, suggests that our value here is also on the low side. The four TFFs therefore have quite different shapes. $E_1(Q^2)$ decreases with $Q^2$ for $h_c\rightarrow\eta_c$, has a minimum for $\psi(2S)\rightarrow\chi_{c0}$, and increases for $\chi_{c0}\rightarrow J/\psi$ and $\eta_c(2S)\rightarrow h_c$. All four are obtained with the same $m_c$ and $\beta$, without any parameter tuned to the transitions.

In the $q^+=0$ frame adopted here one has $q^2=q^+q^--\mathbf{q}_\perp^2=-\mathbf{q}_\perp^2\le0$, so that $Q^2\equiv-q^2=\mathbf{q}_\perp^2\ge0$ and the calculation accesses the space-like region only. The physical radiative decay corresponds to a real photon, $q^2=0$, i.e. the end point $Q^2=0$ of that region, which is where the coupling constant $\mathcal{G}_{SV\gamma}=E_1(0)$ is evaluated. The time-like region relevant to the decay is bounded, $0<q^2\le q^2_{max}=(M_i-M_f)^2$, and is not reached in this frame, so we do not continue the form factor into it. For the four transitions of Fig. \ref{realtmds6} this interval is in any case very small, $q^2_{max}=0.073$ GeV$^2$ for $\psi(2S)\rightarrow\chi_{c0}(1P)$, $0.294$ GeV$^2$ for $h_c(1P)\rightarrow\eta_c(1S)$, $0.101$ GeV$^2$ for $\chi_{c0}(1P)\rightarrow J/\psi(1S)$ and $0.013$ GeV$^2$ for $\eta_c(2S)\rightarrow h_c(1P)$, so that the recoil correction between $q^2=0$ and $q^2=q^2_{max}$ is small in the heavy-flavour sector. For $\psi(2S)\rightarrow\chi_{c0}(1P)$, however, the form factor varies rapidly with $Q^2$ near the origin, so that the recoil correction, while small, is not entirely negligible there. It is important to mention here that till now, there is no experimental data available for TFFs of these decay processes.

From the TFFs at $Q^2=0$, we have calculated the radiative decay widths of each transition and have presented them in Table \ref{tab:my_label22}. The decay width of the $\chi_{c0}(1P)\rightarrow J/\psi(1S)$ transition is found to be $118.3$ keV, to be compared with $151$ keV from the PDG \cite{ParticleDataGroup:2022pth} and $162$ keV from BLFQ \cite{Wang:2023nhb}. For the $\psi(2S)\rightarrow\chi_{c0}(1P)$ transition we obtain $15.8$ keV against $28.8$ keV from the PDG. This is the channel in which the present treatment is least successful. The reason is structural rather than accidental, in that the $2S$--$1P$ dipole overlap lies close to a zero crossing, so the result is strongly sensitive to the harmonic scale and a single universal $\beta$ does not suffice. When compared to the other decays, the $h_c(1P)\rightarrow\eta_c(1S)$ transition has the largest decay width, $534.5$ keV against the PDG value of $445$ keV. For the $\eta_c(2S)\rightarrow h_c(1P)$ transition, for which no experimental data are available, we obtain $10.7$ keV, to be compared with $26.4$ keV from BLFQ and $36$ keV from the GI model \cite{Barnes:2005pb}. This channel shares the node sensitivity of the $2S$--$1P$ overlap. For the sake of  completeness, we have also studied the branching ratio of each decay and the calculated values have been presented in Table \ref{Branching ratio}. The branching ratio of the $h_c\rightarrow\eta_c(1S)+\gamma$ decay is very high, $68.5\%$, as compared with the PDG value \cite{ParticleDataGroup:2022pth} of $60\pm4\%$. This indicates a dominant contribution to the total decay width of the $h_c(1P)$ particle. On the other hand, the branching ratio of the $\eta_c(2S)\rightarrow h_c(1P)+\gamma$ decay comes out to be $0.09\%$, which is very low as compared to the other decays. This clearly indicates  that $\eta_c(2S)$ has very low probability to decay through the  electric dipole radiative transition.

\begin{table}
    \centering
    \begin{tabular}{|c|c|c|c|c|}
    \hline
         &$\chi_{c0}(1P)\rightarrow J/\psi(1S)$ & $\psi(2S)\rightarrow\chi_{c0}(1P)$ & $h_c(1P)\rightarrow\eta_c(1S)$ & $\eta_c(2S)\rightarrow h_c(1P)$  \\\hline
         Our result&118. 3 & 15. 8 & 534. 5 & 10. 7  \\\hline
        PDG \cite{ParticleDataGroup:2022pth} & 151&28. 8 &445  & -  \\\hline
         BESIII \cite{BESIII:2017gcu, BESIII:2022tfo}& 27& 27. 6 &445  & -  \\\hline
         CLEO \cite{CLEO:2004cbu, CLEO:2005efp, CLEO:2008ero, CLEO:2005vqq}& 216& 27. 1 & 437 & -  \\\hline
         CBAL'86 \cite{Gaiser:1985ix}&- & 29. 1 & - & -  \\\hline
         CNTR'77 \cite{PhysRevLett.38.1324}& -& 21 & - & -  \\\hline
         MRK1'76 \cite{PhysRevLett.37.1596}& -& 22 & - & -  \\\hline
         BLFQ \cite{Wang:2023nhb}& 162& 22 & 489 & 26. 4  \\\hline
         Chen'11 \cite{Chen:2011kpa}&85 & - & 234 & -  \\\hline
         Dudek'09 \cite{Dudek:2009kk}&199 & 26 & - & -  \\\hline
         Dudek'06 \cite{Dudek:2006ej}& 232& - & 601 & -  \\\hline
         Delaney'23 \cite{Delaney:2023fsc}& 204& - & - & -  \\\hline
         Li'21 \cite{li2021radiative}& & 180 & - & -  \\\hline
         Becirevic \cite{Becirevic:2012dc}&- & - & 720 & -  \\\hline
         GI model \cite{Barnes:2005pb}& 114& 26 & 352 & 36  \\\hline
        NR \cite{Barnes:2005pb} &- &- &- &49\\\hline
         SP \cite{PhysRevD.95.034026} &- &- &- &52\\\hline
         LP \cite{PhysRevD.95.034026} &141 &27 &361 &49\\\hline
         SNR$_0$ \cite{Li:2009zu} &167 &74 &764 &146\\\hline
         SNR$_1$ \cite{Li:2009zu} &117 &25 &323 &104\\\hline
         RQM \cite{Ebert:2002pp} &121 &26.3 &560 &41\\\hline
         CCQM \cite{Issadykov:2025}&- & - & 304 & -  \\\hline
    \end{tabular}

    \caption{Decay widths (in units of keV) of the transitions along with experimental results \cite{ParticleDataGroup:2022pth,BESIII:2017gcu, BESIII:2022tfo,CLEO:2004cbu, CLEO:2005efp, CLEO:2008ero, CLEO:2005vqq,Gaiser:1985ix,PhysRevLett.38.1324,PhysRevLett.37.1596}, lattice QCD results \cite{Chen:2011kpa,Dudek:2009kk,Dudek:2006ej,Delaney:2023fsc,li2021radiative,Becirevic:2012dc} and other model results \cite{Barnes:2005pb,PhysRevD.95.034026}, \cite{Li:2009zu,Ebert:2002pp,Issadykov:2025}. NR and GI denote the non-relativistic and Godfrey-Isgur potential models, LP and SP the linear and screened potential models, SNR$_{0,1}$ the screened non-relativistic model at zeroth and first order in relativistic corrections, RQM the relativistic quark model and CCQM the covariant confined quark model.}
    \label{tab:my_label22}
\end{table}
\begin{table}
    \centering
    \begin{tabular}{|c|c|c|c|c|}
    \hline
          & $\psi(2S)\rightarrow\chi_{c0}(1P)$ & $h_c(1P)\rightarrow\eta_c(1S)$ & $\eta_c(2S)\rightarrow h_c(1P)$ &$\chi_{c0}(1P)\rightarrow J/\psi(1S)$ \\\hline
        Our result & 5. 38\% & 68. 5\% & 0. 09\% & 1. 13\%\\\hline
        PDG \cite{ParticleDataGroup:2022pth} & 9. 7\% & $60\pm 4\%$ & - & $1. 41\pm 0. 09\%$\\\hline
    \end{tabular}
    \caption{The branching ratios of the calculated decays compared with PDG \cite{ParticleDataGroup:2022pth} data.  }
    \label{Branching ratio}
\end{table}

\section{Summary}
In this work, we have studied the radially excited states of charmonia of pseudo-scalar and vector mesons in the light-front quark model. To begin with, we have predicted the decay constant of $2S$ and $1P$ wave charmonia and the results are comparable to the PDG data. Further, we have calculated the $E_1(Q^2)$ radiative transitions between the $P$- and $S$-wave charmonia, using a decomposition of the electric dipole matrix element into its Lorentz structures and $P$-wave spin-orbit wave functions constructed from the Melosh transformation of the $|(LS)JJ_z\rangle$ states. The calculated transition form factors (TFFs) are found to be in good agreement with the PDG data, lattice simulations and basis light-front quantization (BLFQ) data. The TFFs of all four transitions have been compared with BLFQ, and those of $h_c(1P)\rightarrow\eta_c(1S)$ and $\chi_{c0}(1P)\rightarrow J/\psi(1S)$ with lattice results from more than one group, which do not always agree among themselves. For $\eta_c(2S)\rightarrow h_c(1P)$ no lattice result exists yet, and our prediction differs noticeably from BLFQ in this channel, which we trace back to the node of the $2S$ wave function. For $h_c(1P)\rightarrow\eta_c(1S)$, a more precise measurement of the total width of $h_c$ would discriminate between the lattice results. For a better understanding of these decays, we have calculated the decay widths and branching ratios at the $Q^2=0$ limit. These result are again in sync with the available experimental, lattice simulation, and theoretical prediction results. For the branching ratio, we have compared it with the PDG data and found it to have similar results. To conclude we can say that overall the LFQM provides reasonable results which are in agreement with experimental and other theoretical predictions. Ongoing BESIII. CLEO, Belle, Belle II, BaBar,  PANDA, and  LHCb experiments will have important implications for the subtle features of the model and will further provide deeper insights on these decays.

\section{Acknowledgement}
H. D.  would like to thank  the Science and Engineering Research Board,  Anusandhan-National Research Foundation,  Government of India under the scheme SERB-POWER Fellowship (Ref No.  SPF/2023/000116) for financial support. 

\section{Reference}

%

\end{document}